\documentstyle[12pt]{article}
\begin{document}
\pagestyle{plain}
\setcounter{page}{1}
\baselineskip16pt

\begin{titlepage}

\begin{flushright}
PUPT-1476\\
hep-ph/9406383
\end{flushright}
\vspace{20 mm}

\begin{center}
{\huge On Hyperfine Splittings of Strange}

\vspace{5mm}

{\huge Baryons in the Skyrme Model}
\end{center}

\vspace{10 mm}

\begin{center}
{\large Karl M.\ Westerberg and Igor R.\ Klebanov}

\vspace{3mm}

Joseph Henry Laboratories\\
Princeton University\\
Princeton, New Jersey 08544

\end{center}

\vspace{2cm}

\begin{center}
{\large Abstract}
\end{center}

We calculate the complete order $1/N$ corrections to
baryon masses in the rigid rotator approach to the
3-flavor Skyrme model.

\vspace{2cm}
\begin{flushleft}
June 1994
\end{flushleft}
\end{titlepage}
\newpage
\renewcommand{\baselinestretch}{1.1}  


%
%
%



\newcommand{\AAA}[3]{A_{#1} A_{#2} A_{#3}}
\newcommand{\commAA}[2]{\left[A_{#1},A_{#2}\right]}
\newcommand{\DK}[1]{D_{#1}K}
\newcommand{\dK}[1]{\partial_{#1}K}
\newcommand{\hc}[1]{{#1}^\dagger}
\newcommand{\leftparens}[1]{\left( #1 \right.}
\newcommand{\mAA}[2]{A_{#1} A_{#2}}
\newcommand{\mtxIIIxIII}[4]{\parens{
   \begin{array}{c|c}
      {\displaystyle \vphantom{\underline{#1}} #1}
         & {\displaystyle \vphantom{\underline{#2}} #2} \\ \hline
      {\displaystyle \vphantom{\sqrt{#3}} #3}
         & {\displaystyle \vphantom{\sqrt{#4}} #4}
   \end{array}
}}
\newcommand{\noparens}[1]{#1}
\newcommand{\parens}[1]{\left( #1 \right)}
\newcommand{\power}[2]{{#1}^{#2}}
\newcommand{\rightparens}[1]{\left. #1 \right)}
\newcommand{\taudot}[1]{#1 \cdot \vec{\tau}}
\newcommand{\trAAA}[3]{tr\parens{\AAA{#1}{#2}{#3}}}
\newcommand{\trAA}[2]{tr\parens{\mAA{#1}{#2}}}

\newcommand{\lagtermI}{
-{E_0}
}

\newcommand{\lagtermII}{
-{\Gamma m_K^2 \hc{D} D}
}

\newcommand{\lagtermIII}{
+\frac{1}{2}i N \parens{\hc{D} \dot{D}-\hc{\dot{D}} D}
}

\newcommand{\lagtermIIII}{
+4 \Phi \hc{\dot{D}} \dot{D}
}

\newcommand{\lagtermV}{
+\frac{1}{2} \Omega \power{\dot{\vec{\alpha}}}{2}
}

\newcommand{\lagtermVI}{
+\frac{2}{3} \Gamma m_K^2 \power{\parens{\hc{D} D}}{2}
}

\newcommand{\lagtermVII}{
-{\frac{1}{3}i N \parens{\hc{D} \dot{D}-\hc{\dot{D}} D}} \hc{D} D
}

\newcommand{\lagtermVIII}{
+2 \Phi \power{\parens{\hc{D} \dot{D}-\hc{\dot{D}} D}}{2}
}

\newcommand{\lagtermVIIII}{
-{\frac{1}{2} \parens{\Omega-\frac{4}{3} \Phi} \power{\parens{\hc{D}
\dot{D}+\hc{\dot{D}} D}}{2}}
}

\newcommand{\lagtermX}{
+2 \parens{\Omega-\frac{4}{3} \Phi} (\hc{D} D) (\hc{\dot{D}} \dot{D})
}

\newcommand{\lagtermXI}{
-{\frac{1}{2} N \hc{D} \,\taudot{\dot{\vec{\alpha}}}\, D}
}

\newcommand{\lagtermXII}{
+i \parens{\Omega-2 \Phi} \parens{\hc{D} \,\taudot{\dot{\vec{\alpha}}}\,
\dot{D}-\hc{\dot{D}} \,\taudot{\dot{\vec{\alpha}}}\, D}
}

\newcommand{\palphatermI}{
+i \parens{\Omega-2 \Phi} \parens{\hc{D} \vec{\tau} \dot{D}-\hc{\dot{D}}
\vec{\tau} D}
}

\newcommand{\palphatermII}{
-{\frac{1}{2} N \hc{D} \vec{\tau} D}
}

\newcommand{\palphatermIII}{
\Omega \dot{\vec{\alpha}}
}

\newcommand{\pchitermI}{
4 \Phi \dot{D}
}

\newcommand{\pchitermII}{
-{\frac{1}{2}i N D}
}

\newcommand{\pchitermIII}{
+\frac{1}{3} N (\hc{D} D) D
}

\newcommand{\pchitermIIII}{
+2 \parens{\Omega-\frac{4}{3} \Phi} (\hc{D} D) \dot{D}
}

\newcommand{\pchitermV}{
-{4 \Phi \parens{\hc{D} \dot{D}-\hc{\dot{D}} D}} D
}

\newcommand{\pchitermVI}{
-\parens{{\Omega}-\frac{4}{3} \Phi} \parens{\hc{D} \dot{D}+\hc{\dot{D}} D} D
}

\newcommand{\pchitermVII}{
-{i \parens{\Omega-2 \Phi} \taudot{\dot{\vec{\alpha}}}\, D}
}

\newcommand{\hamAtermI}{
E_0
}

\newcommand{\hamAtermII}{
+\parens{\Gamma m_K^2+\frac{\power{N}{2}}{16 \Phi}} \hc{D} D
}

\newcommand{\hamAtermIII}{
-i {\frac{N}{8 \Phi} \parens{\hc{D} \Pi-\hc{\Pi} D}}
}

\newcommand{\hamAtermIIII}{
+\frac{1}{4 \Phi} \hc{\Pi} \Pi
}

\newcommand{\hamAtermV}{
+\frac{1}{2 \Omega} \power{\vec{J}_{ud}}{2}
}

\newcommand{\hamAtermVI}{
+\parens{\frac{\power{N}{2}}{12 \Phi}-{\frac{2}{3} \Gamma m_K^2}}
\power{\parens{\hc{D} D}}{2}
}

\newcommand{\hamAtermVII}{
-i {\frac{N}{8 \Phi} \parens{\hc{D} \Pi-\hc{\Pi} D}} \hc{D} D
}

\newcommand{\hamAtermVIII}{
-{\frac{1}{8 \Phi} \power{\parens{\hc{D} \Pi-\hc{\Pi} D}}{2}}
}

\newcommand{\hamAtermVIIII}{
+\parens{\frac{1}{12 \Phi}-{\frac{1}{8 \Omega}}} \power{\parens{\hc{D}
\Pi+\hc{\Pi} D}}{2}
}

\newcommand{\hamAtermX}{
+\parens{\frac{1}{2 \Omega}-\frac{1}{3 \Phi}} (\hc{D} D) (\hc{\Pi} \Pi)
}

\newcommand{\hamAtermXI}{
+\frac{N}{4 \Phi} \hc{D} \,\taudot{\vec{J}_{ud}}\, D
}

\newcommand{\hamAtermXII}{
+i\parens{\frac{1}{2 \Omega}-\frac{1}{4 \Phi}} \parens{\hc{D}
\,\taudot{\vec{J}_{ud}}\, \Pi-\hc{\Pi} \,\taudot{\vec{J}_{ud}}\, D}
}

\newcommand{\hamBtermI}{
E_0
}

\newcommand{\hamBtermII}{
+\frac{N}{8 \Phi} (\mu - 1) \hc{a} a
}

\newcommand{\hamBtermIII}{
+\frac{1}{2 \Omega} \power{\vec{J}_{ud}}{2}
}

\newcommand{\hamBtermIIII}{
+\parens{\frac{1}{8 \Omega}-\frac{1}{8 \Phi \mu^2} (\mu - 1)}
\power{\parens{\hc{a} a}}{2}
}

\newcommand{\hamBtermV}{
+\parens{\frac{1}{2 \Omega}-\frac{1}{4 \Phi \mu} (\mu - 1)} \hc{a}
\,\taudot{\vec{J}_{ud}}\, a
}

\renewcommand{\epsilon}{\varepsilon}
\newcommand{\SU}[1]{\mbox{${\rm SU}(#1)$}}
\newcommand{\WZ}{{\rm WZ}}
\newcommand{\QCD}{{\rm QCD}}
\newcommand{\disk}{{\rm D}}
\newcommand{\grad}{\nabla}
\newcommand{\MeV}{{\rm MeV}}
\newcommand{\rightvbar}[1]{\left. #1 \right|}
\newcommand{\tr}{\mathop{\rm tr}}
\newcommand{\sym}{\mathop{\rm sym}}
\newcommand{\id}{1}
\newcommand{\mtxIIxI}[2]{\parens{
   \begin{array}{c}
      #1 \\
      #2
   \end{array}
}}
\newcommand{\mtxIIxII}[4]{\parens{
   \begin{array}{cc}
      #1 & #2 \\
      #3 & #4
   \end{array}
}}

\newcommand{\sA}{{\cal A}}
\newcommand{\sB}{{\cal B}}
\newcommand{\sD}{{\cal D}}
\newcommand{\sL}{{\cal L}}
\newcommand{\sM}{{\cal M}}

\newcommand{\va}{\vec{a}}
\newcommand{\vb}{\vec{b}}
\newcommand{\vn}{\vec{n}}
\newcommand{\vv}{\vec{v}}
\newcommand{\vw}{\vec{w}}
\newcommand{\vx}{\vec{x}}
\newcommand{\vJ}{\vec{J}}
\newcommand{\valpha}{\vec{\alpha}}
\newcommand{\dalpha}{\dot{\valpha}}
\newcommand{\vtau}{\vec{\tau}}

\newcommand{\bc}{\bar{c}}
\newcommand{\bs}{\bar{s}}
\newcommand{\bomega}{\bar{\omega}}

\newcommand{\unitr}{\widehat{r}}
\newcommand{\unittheta}{\widehat{\theta}}
\newcommand{\unitphi}{\widehat{\phi}}

\newcommand{\dr}{{\rm\bf d}r}
\newcommand{\dtheta}{{\rm\bf d}\theta}
\newcommand{\dphi}{{\rm\bf d}\phi}
\newcommand{\dt}{{\rm\bf d}t}

\section{Introduction}

After years of neglect, the Skyrme model \cite{ref1} made a
comeback in the 1980's \cite{ref2,ref4} as a phenomenological low
energy model of hadrons.  In its simplest form,
the Skyrme model is represented by a chiral lagrangian with the
pseudoscalar
mesons as the fundamental fields.  The baryons are formed from meson
solitons with non-zero winding number.

Using the simplest chiral lagrangian which supports baryons of finite
size and energy, the low energy properties of the nucleons
and deltas can be
derived from pion properties and agree with experiment to within 30
percent \cite{ref12,ref13,ref14}.  This is remarkable because the
lagrangian is truncated to fourth order in derivatives.

Difficulties arise in trying to generalize the Skyrme model to
include the strange quark.  Because the strange quark mass is
quite large, the \SU{3} flavor symmetry is significantly
broken.  Although the baryon quantum numbers are predicted correctly
by the \SU{3} Skyrme model \cite{ref7,ref8,ref9,ref10,ref11}, early
attempts at quantitative calculations yield baryon masses which are
way off from experiment, even when the strange quark mass is included
using first order perturbation theory \cite{ref9,ref10,ref11}.

It is hoped that this problem has more to do with the relatively
large strange
quark mass than with defects in the Skyrme model itself
\cite{bs1985,bs1988,ya,ref17,Kaplan&Klebanov}.
The Skyrme lagrangian
may be treated under the rigid rotator approximation with the number of
colors $N$ fixed to 3, but this does not work well.
An alternative approach was proposed
where hyperons are treated as kaon-soliton bound states
\cite{bs1985,bs1988}.  In this scheme, a $1/N$ expansion
for hyperon properties is constructed where in principle
the strange quark mass
may be included exactly at each order. This approach has yielded
good agreement with many observed properties of strange
baryons \cite{bs1985,bs1988,bs1988b,bs1992,Klebanov}.

In fact,
the $1/N$ expansion \cite{ref5}
is an essential part of the Skyrme model because
the connection between Skyrme model and QCD is based on the large $N$
limit \cite{ref3,ref4}.
It is important that the spectrum of low strangeness, low isospin
baryons makes sense in the large $N$ limit.  In other words, the
structure of the multiplets should not change for
increasing values of $N$.  For two flavors this is not a problem
since the \SU{2} multiplets for given spin contain the same number
of baryons for arbitrary $N$.  More precisely, for a given value of
$N$, the allowed values of isospin are given by integral or
half-odd-integral values, depending on whether $N$ is odd or even,
ranging from
$0$ or $\frac12$ up to $N/2$.
Since spin and isospin are the same under the $J=I$ rule, the
low-lying spin-flavor multiplets are independent of $N$.
The situation is
more complicated for three flavors, because the allowed
spin-flavor multiplets
grow in size as $N$ is increased. In fact, even the smallest
multiplets contain baryons with strangeness up to $\sim N$.
The strange quark mass breaks the \SU{3} flavor symmetry and
large strangeness baryons have much larger masses than low
strangeness baryons. Thus, in the large $N$ limit the mass splittings
within each multiplet are large, and in some sense the flavor symmetry
is badly broken. It is convenient, therefore, to take the large
$N$ limit for baryons of fixed strangeness. The
low-lying spin-isospin quantum
numbers of such baryons are independent of $N$, removing any
conceptual problems. This is precisely the route taken in the bound
state approach.

In calculating the baryon masses using the bound state approach, the
hamiltonian is expanded to order $1/N$ and compared with the quark
model where phenomenological
magnetic moment interactions are included.
Calculations of this kind are carried out in
\cite{bs1985,bs1988,bs1988b,bs1992,Klebanov},
where the hamiltonian is treated
exactly to order $N^0$ and some  $\sim 1/N$ terms are taken into
account. However, the
strange-strange interactions, embodied in the terms in
the chiral lagrangian that are quartic in the kaon field,
were not included because of their complexity.
Nevertheless, the model is remarkably successful because the
calculated ratio of the strange-light to light-light interaction
strengths, denoted by $c$, turns out to be close to the empirical value.
It is also important to calculate the parameter $\bc$,
the ratio of the strange-strange to light-light interaction strengths.
Unlike $c$, $\bc$ is sensitive to the terms quartic in the kaon field,
and our goal in this paper is to estimate their effect.

Inclusion of the quartic terms in the full bound state approach
is a difficult task.\footnote{See, however, ref.
\cite{bs1994} for some recent progress.}
For this reason, we will use the rigid rotator Skyrmion as
a testing ground for our methods. Following ref.
\cite{Kaplan&Klebanov}, we develop a
$1/N$ expansion for the 3-flavor rigid rotator by
treating the deviations into the strange directions as perturbations.
This model bears a strong resemblance to the bound state approach,
but is much simpler technically. Essentially, the dynamics of the
kaon field is replaced by that of its most tightly bound mode.
The price we pay is that the wave function of this mode is only
an approximation to what it is in the full
bound state approach, and this approximation becomes
cruder with increasing kaon mass.
Thus some numerical accuracy is
lost, but the calculations become much more manageable
and can be carried out analytically.

A calculation of the hyperfine splittings,
neglecting the strange-strange interaction terms,
has been carried out in \cite{Kaplan&Klebanov}.
There it was found that the perturbative treatment of the strange
quark mass breaks down, and that it has to be included exactly.
In this paper we complete the
calculation of the rigid rotator skyrmion masses
to order $1/N$ by including the strange-strange
interactions.  Although the value of $c$ is unaffected by these
additional terms, the value of $\bc$ is seen to improve vastly over
the partial calculations.  The purpose of completing the rigid
rotator calculation is twofold: (1) hopefully the improvement we
observe in completing the rigid rotator calculation will carry over
to the bound state approach, and (2) the complete rigid rotator
calculation can be used to gain intuition about the complete bound
state calculation.

This paper is organized as follows.  In section 2, we introduce the
Skyrme action.  In section 3 we discuss the rigid rotator
approximation and express the Skyrme action in terms of the rigid
rotator excitations.  In section 4, we carry out the $1/N$ expansion.
Finally, in section 5 we quantize the resulting
lagrangian and calculate $c$ and $\bc$.

\section{The Skyrme model}

The Skyrme lagrangian is given by
\begin{eqnarray}
\sL &=& \frac{f_\pi^2}{16} \tr(\partial_\mu \hc{U} \partial^\mu U)
   + \frac{1}{32e^2} \tr[\partial_\mu U \hc{U},\partial_\nu U \hc{U}]^2
   + \frac{f_\pi^2}{8} \tr \sM(U + \hc{U} - 2) \nonumber \\
 &=& -\frac{f_\pi^2}{16} \tr M_\mu M^\mu
   + \frac{1}{32e^2} \tr[M_\mu,M_\nu]^2
   + \frac{f_\pi^2}{8} \tr \sM(U + \hc{U} - 2) \label{eqn1}
\end{eqnarray}
where $U(\vx,t)\in \SU{3}$. $\sM$ is proportional to the quark
mass matrix and, if we neglect the $u$ and $d$ masses, is given
by
$$ \sM= \mtxIIIxIII{0}{0}{\hc{0}}{m_K^2} $$
where, in general, we write $3\times3$ matrices in
the partitioned form
$$\mtxIIIxIII{2\times2}{2\times1}{1\times2}{1\times1}$$
$M_\mu$ is defined as
$$M_\mu := \partial_\mu U \hc{U}$$
If the quark masses are zero, the lagrangian is
invariant under independent left and right global
\SU{3} rotations of $U$
$$U \mapsto \sA U \hc{\sB}$$
since under this transformation
$$M_\mu \mapsto \sA M_\mu \hc{\sA}$$
Eq.~\ref{eqn1} is the simplest lagrangian invariant
under this symmetry group for zero quark masses which also allows for
classically stable baryons of finite radius and mass.  Without the
commutator term the baryon will collapse, but there is no a priori
reason to exclude other higher derivative
terms.  With a non-zero strange quark mass
the \SU{3} flavor symmetry is broken to the \SU{2}
involving the massless $u$ and $d$
quarks.

There is another term which we must include in the action.
It is called
the Wess-Zumino term \cite{ref6}
and is written in non-local form as an integral
over a five-dimensional disk with spacetime as its boundary:
\begin{equation} \label{eqn3}
S_\WZ = -\frac{iN}{240\pi^2} \int_{\disk} d^5 \vx\,
   \epsilon^{\mu\nu\alpha\beta\gamma}
   \tr(M_\mu M_\nu M_\alpha M_\beta M_\gamma)
\end{equation}
where $U(\vx,t)$ is continuously extended to the disk (with our convention,
$\epsilon^{01234}=1$).
In our treatment,
the Wess-Zumino term is responsible for preventing anti-strange
quarks from appearing in baryons.  To obtain agreement with QCD, $N$
in equation~\ref{eqn3} is taken to be the number of colors
\cite{ref4}.

\section{Rigid rotator approximation}

In this paper we will be concerned with the low-lying baryon
states. We will construct their approximate
description\footnote{Due to the significant \SU{3} breaking,
the bound state approach is, at the end, necessary to improve on
this approximation.} by quantizing the
time-dependent rotations of the $B=1$ Skyrme soliton
$$U(\vx,t) = \sA(t) U_0(\vx) \hc{\sA(t)}$$
We choose the ``hedgehog'' anzatz
$$U_0(\vx)
   = \mtxIIIxIII{e^{iF(\psi)\unitr \cdot \vtau}}{0}{\hc{0}}{1}$$
where $\psi$ is defined by $|\vx|=e^\psi/(e f_\pi)$.
The radial profile function $F(\psi)$ satisfies the boundary conditions
$$F(-\infty) = \pi \qquad F(+\infty) = 0$$
and is determined classically.
Note that the static soliton $U_0$ lies entirely in
the upper-left $2\times 2$ corner.  This is reasonable because the
$N$'s and $\Delta$'s are built out of light quarks.

We will now include the effect of rotation in flavor space.  Calculation
yields
\begin{eqnarray*}
\partial_\mu U &=& \sA\parens{
   \partial_\mu U_0 + [\hc{\sA}\partial_\mu \sA,U_0]
}\hc{\sA} \\
 &=:& \sA\, d_\mu U_0 \,\hc{\sA}
\end{eqnarray*}
Since the action involves only traces of $U$'s and their derivatives,
it follows that the rotation ``gauges'' the action in the following way,
\begin{eqnarray*}
 U &\mapsto& U_0 \\
 \partial_\mu &\mapsto& d_\mu \\
 \sM &\mapsto& \hc{\sA}\sM\sA
\end{eqnarray*}
The action is given by the sum of eqs. \ref{eqn1}
and~\ref{eqn3} with the redefinition above.
Thus the effect of the time-dependent rotation is now hidden in
the definition of $\sM$ and in the covariant derivative $d_\mu$
given by
$$d_\mu U = \partial_\mu U + [\hc{\sA}\partial_\mu \sA,U]$$

Since $\sA$ belongs to \SU{3} it follows that
$\hc{\sA}\dot{\sA}$ is anti-hermitian and traceless, and so it can be
expressed as a linear combination of $i\lambda_a$:
\begin{equation} \label{eqn10}
\hc{\sA}\dot{\sA} = (ef_\pi) i v^a \lambda_a
   = ief_\pi \mtxIIIxIII{\vv\cdot\vtau + \nu\id}{V}{\hc{V}}{-2\nu}
\end{equation}
where
$$\vv = (v^1,v^2,v^3) \qquad
  V = \mtxIIxI{v^4 - iv^5}{v^6 - iv^7} \qquad
  \nu = v^8 / \sqrt3$$

We wish to express $L := \int d^3x\,\sL$ in terms
of $F(\psi)$, $\vv$, $V$ and $\nu$.  First we compute $M_\mu$ which
can be similarly expressed in terms of \SU{3} generators
\begin{equation} \label{eqn4}
M_\mu = (ef_\pi) i w_\mu^a \lambda_a
   = ief_\pi \mtxIIIxIII{\vw_\mu \cdot\vtau + \omega_\mu \id}{
       W_\mu}{\hc{W_\mu}}{-2\omega_\mu}
\end{equation}
Calculation yields
\begin{eqnarray*}
d_r U = \partial_r U &=&
  ief_\pi e^{-\psi}F'(\psi) \mtxIIIxIII{\unitr\cdot\vtau\, U}{0}{\hc{0}}{0} \\
d_\theta U = \partial_\theta U &=&
  ief_\pi e^{-\psi}\sin F \mtxIIIxIII{\unittheta\cdot\vtau}{0}{\hc{0}}{0} \\
d_\phi U = \partial_\phi U &=&
  ief_\pi e^{-\psi}\sin F \mtxIIIxIII{\unitphi\cdot\vtau}{0}{\hc{0}}{0} \\
d_t U = [\hc{\sA}\dot{\sA},U] &=&
  ief_\pi \mtxIIIxIII{2\sin F\, (\unitr\times\vv)\cdot\vtau}{
     (\id-U)V}{\hc{V}(U-\id)}{0}
\end{eqnarray*}
For an arbitrary vector $\vn$ perpendicular to $\unitr$, we find that
$$(\vn\cdot\vtau)\hc{U} = (\vn \cos F + \vn\times\unitr\, \sin F)\cdot\vtau
   = \vn_F \cdot \vtau$$
where $\vn_F$ is $\vn$ rotated through angle $F$ around $\unitr$.  With
the help of this identity, further calculation yields
\begin{eqnarray*}
\omega_\mu &=& 0 \\
W_i &=& 0 \\
W_t &=& (\id - U)V \\
\vw_t &=& 2\sin F\, (\unitr\times\vv)_F \\
\vw_r &=& e^{-\psi}F'(\psi) \unitr \\
\vw_\theta &=& e^{-\psi}\sin F\, \unittheta_F \\
\vw_\phi &=& e^{-\psi}\sin F\, \unitphi_F
\end{eqnarray*}

If we ignore the mass term for now, we can
compute $\sL$ in terms of $W_t$ and $\vw_\mu$, which in turn can
be expressed in terms of $F$, $\vv$ and $V$. The Wess-Zumino
term can also be included (see, for instance, ref. \cite{Balachandran}).
 After integration, we
find
\begin{equation} \label{eqn6}
L = -E_0 + 2(ef_\pi)^2\, \Omega \vv^2 + 2(ef_\pi)^2\, \Phi \hc{V}V
    + (ef_\pi)\, N\nu
\end{equation}
where the classical ground state energy $E_0$, and the two moments
of inertia $\Omega$ and $\Phi$ are given in terms of $F(\psi)$ as
follows
\begin{eqnarray*}
E_0 &=& \pi\frac{f_\pi}{e} \int_{-\infty}^{+\infty} e^{3\psi}d\psi\,
   \parens{\frac12 e^{-2\psi}(F'^2 + 2\sin^2 F)
      + 2e^{-4\psi}\sin^2 F\, (2F'^2 + \sin^2 F) } \\
\Omega &=& \pi\frac{1}{e^3 f_\pi} \int_{-\infty}^{+\infty} e^{3\psi}d\psi\,
   \sin^2 F \parens{\frac23 + \frac83 e^{-2\psi}(F'^2 + \sin^2 F)} \\
\Phi &=& \pi\frac{1}{e^3 f_\pi} \int_{-\infty}^{+\infty} e^{3\psi}d\psi\,
   \frac{1-\cos F}{2} \parens{1+e^{-2\psi}(F'^2 + 2\sin^2 F)}
\end{eqnarray*}
The profile function is determined classically by minimizing
$E_0$ with respect to $F(\psi)$ with fixed boundary conditions.
Numerical integration yields
$$E_0 = 36.4 \frac{f_\pi}{e} \qquad \Omega = 99.1 \frac{1}{e^3 f_\pi}
   \qquad \Phi = 37.8 \frac{1}{e^3 f_\pi}$$

If we write $\sA(t)$ in local coordinates
$$\sA(t) = \sA_0 e^{\frac12 ia^\ell \lambda_\ell}
  \approx \sA_0 (\id + \frac12 ia^\ell \lambda_\ell + \ldots)$$
then $\hc{\sA}\dot{\sA} \approx \frac12 i \dot{a}^\ell \lambda_\ell$
from which we
identify $\dot{a}^\ell = 2ef_\pi v^\ell$.  Then equation~\ref{eqn6}
can be reexpressed as
$$L = -E_0 + \frac12 \Omega \sum_{j=1}^{3} \dot{a}_j^2
      + \frac12 \Phi \sum_{\ell=4}^{7} \dot{a}_\ell^2
      + \frac{N}{2\sqrt3} \dot{a}_8$$
which is a convenient form for quantization.

\section{The $1/N$ expansion}

So far the number of colors has yet to appear, except in the
Wess-Zumino term where it appears explicitly.  QCD without quark
masses has but one coupling constant, which upon renormalization is
converted into a mass scale $\Lambda_\QCD$ and disappears.
Traditionally, the 3-flavor Skyrme model was handled by applying
perturbation theory in $m_s/\Lambda_\QCD$,
but it is shown in \cite{ref9,ref10,ref11} that this does
not work.  Rather than do this, we introduce $1/N$ as the expansion
parameter \cite{ref5,ref3} and treat the deviations of the collective
coordinate wave functions into the strange
directions perturbatively \cite{Kaplan&Klebanov}.

To separate the \SU{2} rotations from the deviations into strange directions,
we write \cite{Kaplan&Klebanov}
$$\sA(t) =
\mtxIIIxIII{A(t)}{0}{\hc{0}}{1}S(t) $$
where $A(t) \in \SU{2}$,
and
$$S(t) = \exp i \sum_{a=4}^{7} d^a \lambda_a
  = \exp i \sD$$
where
$$\sD = \mtxIIIxIII{0}{\sqrt 2 D}{\sqrt 2\hc{D}}{0}\ ,\qquad\qquad
   D = \frac{1}{\sqrt2} \mtxIIxI{d^4 - id^5}{d^6 - id^7}$$
It is convenient to introduce the angular velocity of the
\SU{2} rotation via
$$\hc{A}\dot{A} = \frac12 i \dalpha\cdot\vtau$$
Our goal will be to express the lagrangian in eq.~\ref{eqn6}
in terms of $D$ and $\dalpha$.

The momentum conjugate to $\valpha$
is $\vJ_{ud}$, which can be interpreted as
the angular momentum carried by the $u$ and $d$
quarks. The \SU{2} rotator quantization yields its equality to
the isospin, $J_{ud} = I$.  The
low-lying states with isospin of order $1$ correspond to angular
velocities of order $1/N$, which is helpful in developing the
$1/N$ expansion.

$S(t)$ corresponds to deviations into the strange direction.  The
order $1$ strangeness baryons correspond to $D$ values of
order $1/\sqrt{N}$.
There are two effects which keep these deviations small.
The strange quark mass is non-zero, so that the higher
strangeness baryons will have a higher mass.  But even if the strange
quark mass were zero, the Wess-Zumino term acts as a magnetic field
whose strength is order $N$, thereby confining
the wave functions of low strangeness baryons.
Our intention is to calculate the baryon spectrum of these low lying
states completely to order $1/N$.  Since the lagrangian is order $N$
to begin with, we need to retain the previously left out
corrections of order $D^4$,
which incorporate the strange-strange interactions.

We start with the systematic expansion of $S = \exp i\sD$.
In fact, $S$ can be expanded to {\em arbitrary\/}
order in $D$ because $\sD$ satisfies
$$\sD^3 = d^2 \sD \qquad d^2 := 2\hc{D}D$$
so that
\begin{eqnarray*}
S &=& \sum_{n=0}^\infty \frac{i^n}{n!} \sD^n \\
 &=& \id + i\parens{\sum_{n=0}^\infty \frac{(-1)^n}{(2n+1)!} d^{2n}}\sD
   - \parens{\sum_{n=0}^\infty \frac{(-1)^n}{(2n+2)!} d^{2n}}\sD^2 \\
 &=& \id + i\sD \frac{\sin(d)}{d} -  \sD^2 \frac{1-\cos(d)}{d^2}
\end{eqnarray*}
Thus, we have
$$S = \mtxIIIxIII{
   \id - \frac{1-\cos(d)}{d^2} \, 2 D\hc{D}
}{
   i \frac{\sin(d)}{d} \, \sqrt2 D
}{
   i \frac{\sin(d)}{d} \, \sqrt2 \hc{D}
}{
   \cos(d)
}$$

First, we take care of the mass term, $\tr \sM(U + \hc{U} -2)$.
Straightforward calculation yields
$$\tr \sM(U + \hc{U} - 2) = -2m_K^2 (1-\cos F) \sin^2(d) $$
Integrating over space and expanding to order $D^4$ we get
$$L_M = -\Gamma m_K^2 (\hc{D}D - \frac23 (\hc{D}D)^2) + \ldots$$
where $\Gamma$ is order $N$ and is given by
$$\Gamma := \frac{4\pi}{e^3 f_\pi} \int_{-\infty}^\infty d\psi\,
   e^{-3\psi}\parens{\frac{1-\cos F}{2}} = 103.2 \frac{1}{e^3 f_\pi}$$

The other terms in the lagrangian, eq.~\ref{eqn6},
depend on $\vv^2$, $\hc{V}V$ and
$\nu$, which are obtained by comparing eq.~\ref{eqn10} with
$$\hc{\sA}\dot{\sA} = \hc{S}\hc{A}(\dot{A}S + A\dot{S})
  = \hc{S}\mtxIIIxIII{\frac12 i \dalpha\cdot\vtau}{0}{\hc{0}}{1}S
    + \hc{S}\dot{S}$$
A useful identity is
$$\vv^2 = \frac12 \tr(\vv\cdot\vtau + \nu\id)^2 - \nu^2$$
In principle we could calculate $\hc{\sA}\dot{\sA}$ systematically
to all orders in $1/N$, but it is {\em much\/} simpler to truncate
to $D^4$ immediately.  Since
$\vv\cdot\vtau + \nu\id$ starts off at order $D^2$, we only need it to
order $D^2$ to compute $\vv^2$ to order $D^4$.  A straightforward
but tedious calculation yields
\begin{eqnarray*}
ef_\pi(\vv\cdot\vtau + \nu\id)
 &=& \frac12 \dalpha\cdot\vtau + i(\dot{D}\hc{D} - D\hc{\dot{D}})
     + \ldots \\
{ef_\pi V  \over\sqrt2}
 &=& \dot{D} + \frac12 i \dalpha\cdot\vtau\, D - \frac13 (\hc{D}D)\dot{D}
     + \frac16(\hc{D}\dot{D} + \hc{\dot{D}}D)D
     - \frac12(\hc{D}\dot{D} - \hc{\dot{D}}D)D + \ldots \\
ef_\pi \nu
 &=& \frac12 i (\hc{D}\dot{D} - \hc{\dot{D}}D)
     - \frac12 \hc{D}\,\dalpha\cdot\vtau\, D
     - \frac13 i(\hc{D}\dot{D} - \hc{\dot{D}}D) \hc{D}D + \ldots
\end{eqnarray*}
from which, after more calculation we conclude
\begin{eqnarray}
L &=&
\lagtermI
\lagtermIIII
\lagtermIII
\lagtermII
  \nonumber \\ && \mbox{}
\lagtermV
\lagtermXII
  \nonumber \\ && \mbox{}
\lagtermXI
\lagtermX
  \nonumber \\ && \mbox{}
\lagtermVIIII
\lagtermVIII
  \nonumber \\ && \mbox{}
\lagtermVII
\lagtermVI
  \label{eqn7}
\end{eqnarray}

\section{Quantizing the lagrangian}

The complete expansion of the lagrangian to order $1/N$ is given
by equation~\ref{eqn7}.  The quantum variables are the local
coordinates of the \SU{2} rotation, $\valpha$, and the deviations
into the strangeness directions
$d^a, a=4,5,6,7$, arranged conveniently in a complex spinor
$$D = \frac{1}{\sqrt2} \mtxIIxI{d^4 - id^5}{d^6 - id^7}$$
The conjugate momenta are given by the $u$-$d$ quark angular
momentum $\vJ_{ud}$ and $\pi_a, a=4,5,6,7$ which can also be
arranged in a complex spinor
$$\Pi = \frac{1}{\sqrt2} \mtxIIxI{\pi^4 - i\pi^5}{\pi^6 - i\pi^7}$$
The momenta are given by the usual formula
$$(J_{ud})_i = \frac{\delta L}{\delta \dot{\alpha}^i} \qquad
  \Pi^\gamma= \rightvbar{\frac{\delta L}{\delta \hc{D}_\gamma}
     }_{D\;\rm fixed}$$
and satisfy the commutation relations
$$[(J_{ud})_i,\alpha^j] = \frac{1}{i} \delta_i^j \qquad
  [\Pi^\gamma,\hc{D}_\beta] = [\hc{\Pi}_\beta,D^\gamma]
     = \frac{1}{i} \delta_\beta^\gamma$$
Calculation gives
\begin{eqnarray}
\vec{J}_{ud} &=&
\palphatermIII
\palphatermI
\palphatermII
   \\
\Pi &=&
\pchitermI
\pchitermII
\pchitermVII
  \nonumber \\ && \mbox{}
\pchitermVI
\pchitermV
  \nonumber \\ && \mbox{}
\pchitermIIII
\pchitermIII
\end{eqnarray}
The hamiltonian is calculated to order $1/N$ by Legendre-transforming
the lagrangian with the aid of a computer.  The result is
\begin{eqnarray}
H &=& \dalpha\cdot\vJ_{ud} + \hc{\Pi}D + \hc{D}\Pi - L \nonumber \\
&=&
\hamAtermI
\hamAtermIIII
\hamAtermIII
\hamAtermII
  \nonumber \\ && \mbox{}
\hamAtermV
\hamAtermXII
  \nonumber \\ && \mbox{}
\hamAtermXI
\hamAtermX
  \nonumber \\ && \mbox{}
\hamAtermVIIII
\hamAtermVIII
  \nonumber \\ && \mbox{}
\hamAtermVII
\hamAtermVI
  \label{eqn8}
\end{eqnarray}

The order $N$ piece of the hamiltonian is simply the classical
ground state energy $E_0$.  The order $1$ piece includes the terms
quadratic in $D$ and $\Pi$, and thus may be diagonalized exactly
using creation and annihilation operators
$$D^\gamma= \frac{1}{\sqrt{N\mu}} (a^\gamma+ (\hc{b})^\gamma) \qquad
  \Pi^\gamma= \frac{\sqrt{N\mu}}{2i} (a^\gamma- (\hc{b})^\gamma)$$
where
$$\mu = \sqrt{1+(m_K/M_0)^2}\ ,
\qquad\qquad M_0 = \frac{N}{4\sqrt{\Gamma\Phi}}$$
The operators $\hc{a}$ ($\hc{b}$) may be thought of as
creation operators for constituent strange quarks (anti-quarks).
Strangeness and the angular momentum of the
strange quarks are given respectively by
$$S = \hc{b}b - \hc{a}a \qquad
  \vJ_s = \frac12 (\hc{a} \vtau a - b \vtau \hc{b})$$
In terms of the creation and annihilation operators, the
normal-ordered hamiltonian to order $1$ is given by
$$H = E_0 + \omega \hc{a}a + \bomega \hc{b}b\ ; \qquad
 \omega = \frac{N}{8\Phi}(\mu - 1)\ , \qquad
 \bomega = \frac{N}{8\Phi}(\mu + 1)\ .$$
Thus, replacing a light quark with a strange quark (anti-quark)
costs energy $\omega$ ($\bomega$).  Note that for
vanishing $m_K$, $\omega$ also vanishes, thereby restoring the
original \SU{3} symmetry (it costs no energy to replace a $u$ or
$d$ quark with an $s$ quark) but that $\bomega$ tends to a
rather large value, $N/4\Phi$. Indeed, baryons containing
strange anti-quarks are exotic states that have not been observed in
nature.  The Wess-Zumino term, which acts as magnetic field in the
$D$ -- $\hc{D}$ plane, breaks the
$s \leftrightarrow \bs$ symmetry.

{}From now on, we consider those states for which $n_{\bs} = 0$.
$b$ annihilates these states, and since we plan to normal-order
the hamiltonian, we may simply drop the contributions
of the $b$-oscillators.  Upon doing so, we find
that the hamiltonian to order $1/N$ is
\begin{eqnarray}
H &=&
\hamBtermI
\hamBtermII
\hamBtermIII
\hamBtermV
  \nonumber \\ && \mbox{}
\hamBtermIIII
  \label{eqn9}
\end{eqnarray}
Using the identity
$$\vJ_s^2 = \frac14 (\hc{a}a)^2+ \frac12 \hc{a}a$$
we may rewrite the hamiltonian as
$$H = E_0 + \omega \hc{a}a + \frac{1}{2\Omega}\parens{
   \vJ_{ud}^2 + 2c \vJ_{ud} \cdot \vJ_s + \bc \vJ_s^2
}$$
Here $\omega$ differs from $\frac{N}{8\Phi}(\mu - 1)$ by a subleading
term of order $1/N$, which we neglect.
The explicit formulae for $c$ and $\bc$ are
$$c = 1 - \frac{\Omega}{2\mu\Phi}(\mu - 1)\ ,\qquad \qquad
  \bc = 1 - \frac{\Omega}{\mu^2 \Phi}(\mu - 1)\ .$$
Let us note that the structure of the hyperfine (order $1/N$) splittings
is the same as in the quark model with phenomenological magnetic moment
interactions,
$$ H_{hf} = {1\over \Omega} \parens{\sum_{i<k} \vec j_i\cdot \vec j_k +
c \sum_{i,K} \vec j_i\cdot \vec j_K +
\bc \sum_{I<K} \vec j_I\cdot \vec j_K}
$$
(the small indices refer to light quarks, while the capital -- to strange
quarks). It is remarkable that in the Skyrme model such
interactions are derived rather than postulated, and that the values of
the parameters are explicitly calculable. Using the fact that
$\vec J_{ud}^2 = I(I+1)$, the hyperfine splittings can be expressed
as
$$\delta M =\frac{1}{2\Omega}\left\{ cJ(J+1) +
(1-c) (I(I+1) - \frac{Y^2}{4}) + (1+\bc - 2c) \frac{Y^2}{4} \right \}
\ .$$

We use the parameters
$$f_\pi = 129 \,\MeV \qquad e = 5.45$$
from the standard fit to $N$ and $\Delta$.
Thus numerical calculation gives us
$$E_0 = 862\,\MeV \qquad
  \Omega^{-1} = 211 \,\MeV \qquad
  \Phi^{-1} = 552 \,\MeV \qquad
  \Gamma^{-1} = 202 \,\MeV$$
which leads to
$$M_0 = 250 \,\MeV \qquad \mu = 2.22$$
We finally conclude that
$$c = .28 \qquad \bc = .35$$

At this point, we can see why first order perturbation theory in
$m_K^2$ fails so badly: the expansion parameter turns out to be
$m_K^2 / M_0^2$ which is almost as large as $4$! Note that the
unexpectedly low mass scale $M_0$ is peculiar to baryons and
does not appear in purely mesonic physics
\cite{Kaplan&Klebanov}. Thus, the Skyrme model
analysis suggests that chiral perturbation theory for strange baryons
is not reliable. We regard this is an interesting and non-trivial
prediction. Although $M_0$
is model-dependent, we believe that our qualitative conclusions are sound.

\section{Conclusions}

A summary of previous calculations of $c$ and $\bc$ is
given in Table~\ref{tab1}.
\begin{table}
\caption{Values of $c$ and $\bc$} \label{tab1}
\begin{center}
\begin{tabular}{|l|c|c|c|} \hline
Source & $c$ & $\bc$ \\ \hline\hline
Experiment
   & $.67$ & $.27$ \\ \hline
Rigid rotator, partial
   & $.28$ & $.08$ \\ \hline
Rigid rotator, complete
   & $.28$ & $.35$ \\ \hline
Bound state, partial
   & $.60$ & $.36$ \\ \hline
Bound state, complete
   & $.60$ & ? \\ \hline
\end{tabular}
\end{center}
\end{table}
In this paper we include all
terms up to order $1/N$, including the strange-strange interactions.
These interactions affect the value of $\bc$, although they do not
affect $c$.
In the quark model one typically uses the relation
$\bc = c^2$,
which is equivalent to expressing the hamiltonian in the form
$$H = E_0 + \omega \hc{a}a + \frac{1}{2\Omega}(\vJ_{ud} + c\vJ_s)^2$$
The same hamiltonian follows from the bound state approach with the
quartic terms in the kaon field neglected. We have explicitly shown
here that the strange-strange interactions
break this relation, thereby improving the value of $\bc$.

Inclusion of the quartic terms in the bound state approach, and calculation of
their effect on $\bc$, is a laborious calculation which we leave
for the future. Judging from the results presented here, we do not
expect the shift in $\bc$ to be too large. As suggested by the recent
results in ref.~\cite{bs1994},
the validity of the bound state approach will undoubtedly be
preserved.

\section*{Acknowledgements}

This work was supported in part by DOE grant DE-AC02-76WRO3072,
the NSF Presidential Young Investigator Award PHY-9157482,
James S. McDonnell Foundation grant No. 91-48,
and an A. P. Sloan Foundation Research Fellowship.

\end{document}